\documentclass[sensors,communication,accept,pdftex,moreauthors]{Definitions/mdpi}
\usepackage{comment} 

\makeatletter
\let\c@lofdepth\relax
\let\c@lotdepth\relax
\makeatother 
\usepackage{subfigure}
\makeatletter
\renewcommand{\@thesubfigure}{\normalsize(\textbf{\alph{subfigure}})}
\makeatother

\usepackage{makecell}
\usepackage{graphicx}
\usepackage{multirow}

\def\tsc#1{\csdef{#1}{\textsc{\lowercase{#1}}\xspace}}
\tsc{WGM}
\tsc{QE}
\tsc{EP}
\tsc{PMS}
\tsc{BEC}
\tsc{DE}
\firstpage{1} 
\pubvolume{1}
\issuenum{1}
\articlenumber{0}
\pubyear{2024}
\copyrightyear{2024}
\externaleditor{Academic Editor: Firstname Lastname\vspace{-12pt}}
\datereceived{30 November 2023} 
\daterevised{3 January 2024} 
\dateaccepted{29 January 2024} 
\datepublished{ } 
\hreflink{https://doi.org/} 

\Title{{\colorbox{white}{Gait} 
 Event Detection and Travel Distance Using Waist-Worn Accelerometers across a Range of Speeds: Automated {Approach}}}

\TitleCitation{Gait Event Detection and Travel Distance Using Waist-Worn Accelerometers across a Range of Speeds: Automated Approach}

\Author{\textcolor{black}{Albara} 
 Ah Ramli $^{1}$\orcidA{}, Xin Liu $^{1}$, Kelly Berndt $^{2}$, Chen-Nee Chuah $^{5}$, Erica Goude $^{2}$, Lynea B. Kaethler $^{2}$,
Amanda Lopez $^{2}$, Alina Nicorici $^{2}$, Corey Owens $^{3}$, David Rodriguez $^{2}$, Jane Wang $^{2}$, Daniel Aranki $^{4}$, Craig M. McDonald $^{2}$ \linebreak  and \textcolor{black}{Erik K. Henricson} 
 $^{2,}$*}

\AuthorNames{Albara Ah Ramli, Xin Liu, Kelly Berndt, Chen-Nee Chuah, Erica Goude, Lynea B. Kaethler, Amanda Lopez, Alina Nicorici, Corey Owens, David Rodriguez, Jane Wang, Daniel Aranki, Craig M. McDonald and Erik K. Henricson}

\AuthorCitation{\textcolor{black}{Ramli,} 
 A.A.; Liu, X.; Berndt, K.; Chen-Nee, C.; Goude, E.; Kaethler, L.B.; Lopez, A.; Nicorici, A.; Owens, C.; Rodriguez, D.; et al.}

\address{%
$^{1}$ \quad Department of Computer Science, School of Engineering, University of California, 1 Shields Ave, \linebreak  Davis, CA 95616, USA; \textcolor{black}{arramli@ucdavis.edu (A.A.R.); xinliu@ucdavis.edu (X.L.)} 
\\
$^{2}$ \quad Department of Physical Medicine and Rehabilitation, School of Medicine, University of California, \linebreak  1 Shields Ave, Davis, CA 95616,  USA; \textcolor{black}{klberndt@ucdavis.edu (K.B.); emgoude@ucdavis.edu (E.G.); 
lbkaethler@ucdavis.edu (L.B.K.); aflopez@ucdavis.edu (A.L.); arnicorici@ucdavis.edu (A.N.); davrodri@ucdavis.edu (D.R.); ajawang@ucdavis.edu (J.W.); cmmcdonald@ucdavis.edu (C.M.M.)}\\
$^{3}$ \quad UC Davis Center for Health and Technology, School of Medicine, University of California Davis, \linebreak  1 Shields Ave, Davis, CA 95616, USA; \textcolor{black}{cowens@ucdavis.edu (C.O.)}\\
$^{4}$ \quad Berkeley School of Information, University of California Berkeley, 1 Shields Ave, Berkeley, CA 94720,  USA; \textcolor{black}{daranki@berkeley.edu (D.A.)}\\
$^{5}$ \quad Department of Electrical and Computer Engineering, School of Engineering, University of California, \linebreak  1 Shields Ave, Davis, CA 95616, USA; \textcolor{black}{chuah@ucdavis.edu (C.C.)}\\
}

\corres{Correspondence: ehenricson@ucdavis.edu}

\abstract{Estimation of temporospatial clinical features of gait (CFs), such as step count and length, step duration, step frequency, gait speed, and distance traveled, is an important component of community-based mobility evaluation using wearable accelerometers.  However, accurate unsupervised computerized measurement of CFs of individuals with Duchenne muscular dystrophy (DMD) who have progressive loss of ambulatory mobility is difficult due to differences in patterns and magnitudes of acceleration across their range of attainable gait velocities. {This paper proposes a novel calibration method. It aims to detect steps, estimate stride lengths, and determine travel distance. The approach involves a combination of clinical observation, machine-learning-based step detection, and regression-based stride length prediction. The method demonstrates high accuracy in children with DMD and typically developing controls (TDs) regardless of the participant’s level of ability.} Fifteen children with DMD and fifteen TDs underwent supervised clinical testing across a range of gait speeds using 10 m or 25 m run/walk (10 MRW, 25 MRW), 100 m run/walk (100 MRW), \mbox{6-min walk} (6 MWT), and free-walk (FW) evaluations while wearing a mobile-phone-based accelerometer at the waist near the body’s center of mass.  Following calibration by a trained clinical evaluator, CFs were extracted from the accelerometer data using a multi-step machine-learning-based process and the results were compared to ground-truth observation data.  Model predictions vs. observed values for step counts, distance traveled, and step length showed a strong correlation (Pearson’s r = $-$0.9929 to 0.9986, \emph{p} < 0.0001). The estimates demonstrated a mean (SD) percentage error of 1.49\% (7.04\%) for step counts, 1.18\% (9.91\%) for distance traveled, and 0.37\% (7.52\%) for step length compared to ground-truth observations for the combined 6 MWT, 100  MRW, and FW tasks. Our study findings indicate that a single waist-worn accelerometer calibrated to an individual’s stride characteristics using our methods accurately measures CFs and estimates travel distances across a common range of gait speeds in both DMD and TD peers.}


\begin{document}
\section{Introduction}

There is a pressing need to overcome challenges in measuring key temporospatial clinical features (CFs) of children with Duchenne muscular dystrophy (DMD)~\cite{duan2021duchenne} in the community using single mobile wearable acceleration sensors.  Gait patterns of people with DMD become progressively atypical with advancing disease, with~reduced stride lengths, cadence, and~gait speed, reduced intensity of accelerations in the vertical and anteroposterior axes of travel, and~reduced vertical accelerations~\cite{sutherland1981pathomechanics}.  These changes lead to alterations in patterns and magnitudes of acceleration that complicate computer-based identification of individual steps, which then impairs researchers’ ability to measure gait features in an unsupervised manner during community-based activities.  Single-sensor signals have been used to detect important gait events such as initial contact (IC) at different walking and running speeds~\cite{zijlstra2003assessment} in laboratory settings, but~methods for combining signals across an individual’s full range of attainable velocities to accurately measure steps and distance traveled in the community have not been described.  Our method uses a combination of clinical observation and machine-learning- and regression-based approaches to identify the weak patient’s novel patterns of acceleration associated with their ambulation at different speeds.

Accelerometers can be more accurate than pedometers at slower walking speeds and in populations with atypical gait patterns, making pedometers less suitable for evaluating physical activity in such populations~\cite{le2003comparison}. Estimating CFs of gait (step length, step duration, step frequency, and~gait speed) is a fundamental step in gait analysis, and~detecting the IC of the heel is crucial for identifying gait events and the beginning of the step cycle. In~a laboratory environment, detecting events and estimating CFs is typically accomplished by measuring ground reaction forces (GRF) and verifying with visual observation. However, using these methods to measure gait events in the community is often~impractical.

Studies have described the potential of using acceleration signals to estimate CFs. Several studies have demonstrated that step length, gait speed, IC, and~incline can be determined from acceleration signals of the lower trunk~\cite{zijlstra2003assessment}. Aminian and colleagues explored the feasibility of using a fully connected artificial neural network (ANN) with accelerometers on the trunk and heel to predict incline and speed based on ten statistical parameters extracted from the raw signal~\cite{aminian1995incline}. The~results revealed that a negative peak in the heel accelerometer signal indicates IC events in each gait cycle (two steps).

Studies comparing accelerometer signals from different body positions at various walking speeds demonstrate that positions near the body’s center of mass (trunk, waist, pelvis, and~sacrum) are suitable for capturing gait events~\cite{khandelwal2017evaluation,kavanagh2006reliability,gonzalez2010real}. In~a study by Zijlstra~et~al., participants walked on a force-transducing treadmill and overground while trunk acceleration data were recorded to estimate step lengths and walking speed. IC events were matched with vertical ground reaction force (GRF) normalized by body weight to anteroposterior acceleration. The~start and end of gait cycles from the GRF corresponded with the time of the peak amplitude value in the anteroposterior acceleration signal~\cite{zijlstra2003assessment}. Further research by Lee~et~al. and Mo~et~al. demonstrated that IC events can be determined from anteroposterior acceleration measured at the pelvis and sacrum~\cite{lee2010use,mo2018accuracy}. They collected accelerometer signals from the pelvis/sacrum and GRF data and matched IC events on anteroposterior acceleration with vertical GRF. Initial contact events on the force plate corresponded with the instant of the positive peak pelvis/sacrum anteroposterior acceleration~\cite{mo2018accuracy}.

{Detecting IC gait events and precisely measuring walking/running distance are vital components of gait analysis, offering invaluable clinical insights. Although~GRF is capable of detecting IC events, its dependence poses limitations in communities without GRF availability. Therefore, there is a pressing need for alternative methods in IC event detection. Additionally, accurate distance measurement is critical in gait analysis, especially when dealing with muscle disorders. Traditional methods such as pedometers and wheel measurements encounter challenges in communities, particularly for long distances and low speeds, as~observed in participants with muscular disorders. Consequently, a~method that ensures both accurate distance estimation and IC event detection is imperative.}

{We present a machine learning (ML)-based method that automates detection of IC events and CFs using raw accelerometer signals obtained from consumer mobile \mbox{devices~\cite{ramli2023walk4me,DMD_PAPER}.} We demonstrate that using a single accelerometer worn close to the body's center of mass is an accurate and reliable approach to estimate CFs and IC events across a typical range of walking speeds. This method can be applied to healthy individuals and those with gait disturbances without the need for GRF measurements.}

\section{Materials and~Methods}

Estimating distance using accelerometer signals is challenging due to inherent quadratic error of accelerometers, which can result in deteriorating estimates even with short integration times and distances. Many methods attempt to estimate distance from accelerometers by integrating acceleration twice with respect to time, despite incorporating error-limiting mechanisms and setting restrictions, which can result in errors due to noise, drift, and~bias~\cite{alvarez2018accelerometry}. We propose an ML-based signal processing method that accurately estimates an individual's distance traveled, step length, and~number of steps across varying walking/running speeds, outperforming the built-in pedometer function on iPhones, which shows the highest error percentage in slow walking speeds~\cite{lee2010use}.

\begin{figure}[H]

\begin{adjustwidth}{-\extralength}{0cm}
\centering\includegraphics[scale=0.245]{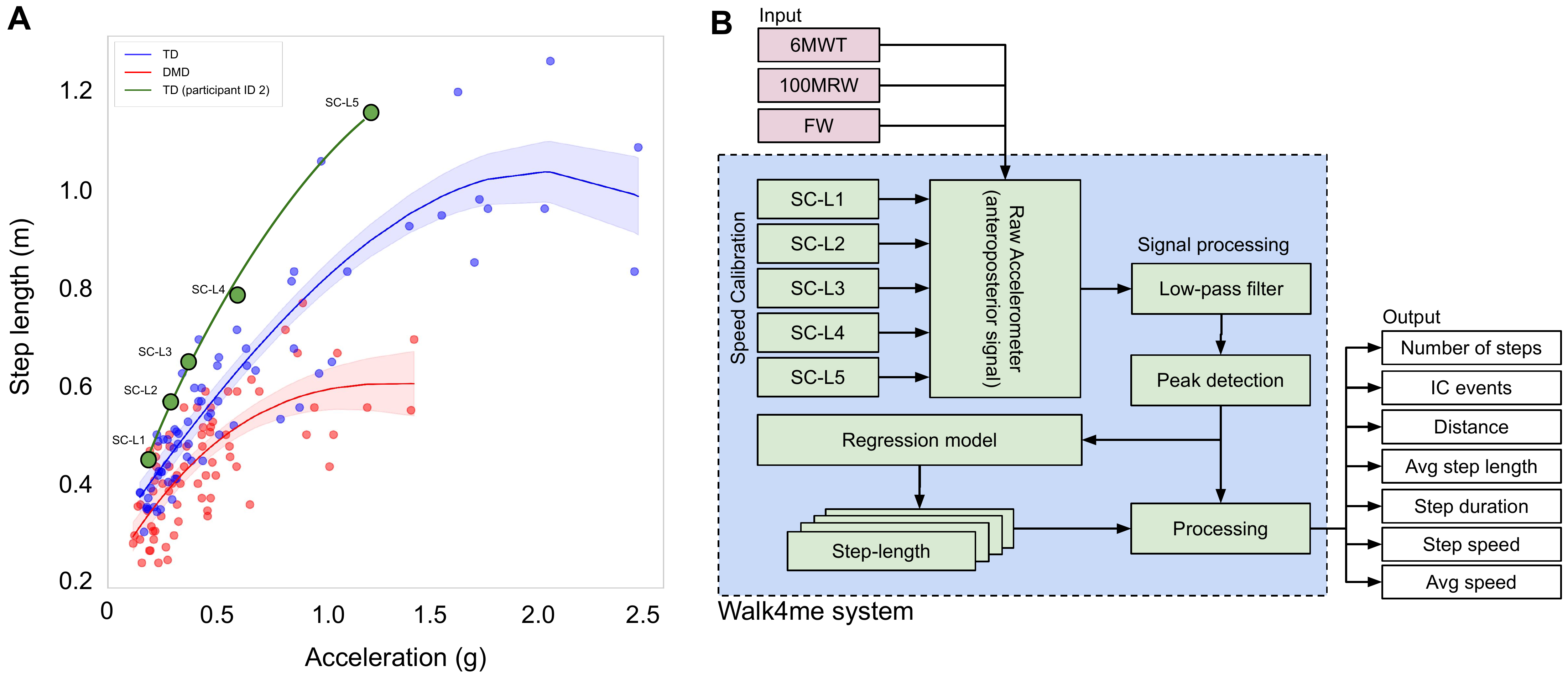}
\end{adjustwidth}
\caption{(\textbf{A}) \textcolor{black}{The} 
 relationship between step length and acceleration of the body's center of mass at various speeds for TD individuals and those with DMD. The~plotted curves depict the regression model, with~the black line representing all participants, the~green line representing TD participants, and~the red line representing DMD participants. (\textbf{B}) The diagram depicts the data flow of our model training and prediction process. In~the training phase, the~model uses five speed calibrations, SC-L1 to SC-L5, with~ground truth to predict the average step length. The~input to our model is the acceleration signal from unseen gait activities (6 MWT, 100 MRW, and~FW).
}\label{reg-pattren_REG_PAPER}\end{figure}

Because different individuals have different walking/running behaviors that affect acceleration, we built a regression model for each individual to estimate distance based on their specific walking/running patterns. We developed a regression model using data from five different speeds (SC-L1 to SC-L5) to map step length to the corresponding anteroposterior acceleration amplitudes using pairs of distance and acceleration values (Figure~\ref{reg-pattren_REG_PAPER}A).  We calculated distance for a single speed by averaging the step distances, while the acceleration was calculated by averaging the maximum values of acceleration in each step (Figure~\ref{reg-pattren_REG_PAPER}B).

\clearpage 
To ensure a fair comparison, we evaluated three sources of estimated data: first, ground-truth data based on video observation of distance traveled and number of steps; second, the~pedometer sensor in the iPhone, which provided estimates of distance and number of steps; and, third, our Walk4Me system~\cite{ramli2023walk4me}, which includes calibration regression models for estimating distance and a signal processing algorithm for measuring number of steps. We estimated the speed, step length, and~frequency as derivatives from the regression and signal~processing.



\subsection{Participants}



Fifteen children with DMD and fifteen TD peers participated in gait speed experiments. The~age of the participants ranged from 3 to 16 years, with~a mean age of 8.6 years and a standard deviation of 3.5. Their body weight ranged from 17.2 to 101 kg, with~a mean weight of 36 kg and a standard deviation of 18.8. Their height ranged from 101.6 cm to \mbox{165.5 cm,} with~a mean height of 129 cm and a standard deviation of 15.8. All participants had at least 6 months of walking experience and were able to perform a 10-m walk/jog/run test in less than 10 s. Participants with DMD had a confirmed clinical diagnosis and were either naïve to glucocorticoid therapy or on a stable regimen for at least three months. Northstar Ambulatory Assessment (NSAA)~\cite{muntoni2022novel} scores for DMD participants ranged from 34 to 8, indicating typical levels of function to clinically apparent moderate mobility limitation (Table \ref{participant_table}). 

\begin{figure}[H]
\includegraphics[scale=0.67]{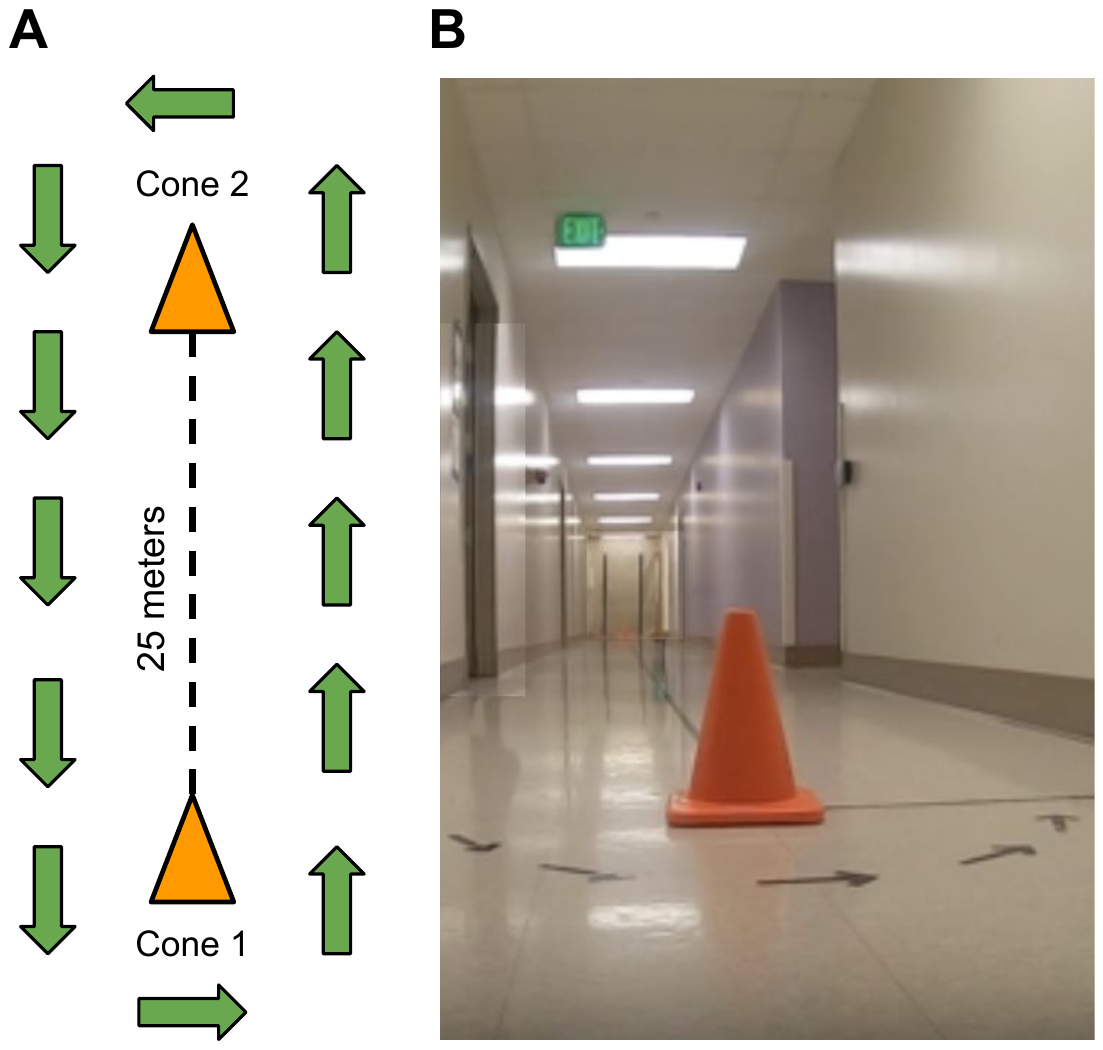}
\caption{(\textbf{A}) Diagram depicting the corridor layout with two cones positioned 25 m apart, guiding walking, running, and~jogging directions. Exception for free-walk (FW), allowing participants the freedom to move within the building. (\textbf{B}) Image showcasing the real-life corridor environment.}\label{Fig0AB}
\end{figure}

The~protocol was reviewed and approved by the Institutional Review Board (IRB) at the University of California, Davis, and~informed consent was obtained from each participant prior to the initiation of study procedures. Measurements were taken at eight different walking/running gait activities, including speed-calibration tests at slow walk to running speeds (SC-L1, SC-L2, SC-L3, SC-L4, and~SC-L5), a~6-min walk test (6 MWT)~\cite{agarwala2020six}, a~100-m fast-walk/jog/run (100 MRW)~\cite{alfano2017100}, and~a free walk (FW). {Participants engaged in walking, jogging, or~running within a 25-m corridor delimited by two cones, as~illustrated in Figure~\ref{Fig0AB}A,B.}

\begin{table}[H]

\caption{Characteristics of the children included in the~study.}\label{participant_table}
\newcolumntype{c}{>{\centering\arraybackslash}X}
\begin{tabularx}{\textwidth}{cccccc}
\toprule
\textbf{ID} & \textbf{Case} & \textbf{Age} & \textbf{Weight} & \textbf{Height} & \textbf{NSAA}\\
   & \textbf{(Status) } & \textbf{(Years)} & \textbf{(kg)} & \textbf{(cm)} & \textbf{(/34)}\\
\midrule
1 & TD & 15 & 101.0 & 165.5 & 34\\ 
2 & TD & 12 & 57.7 & 155.6 & 34\\ 
3 & TD & 11 & 46.4 & 146.0 & 34\\ 
4 & TD & 9 & 41.8 & 132.9 & 34\\ 
5 & TD & 8 & 29.6 & 126.5 & 34\\ 
6 & TD & 8 & 27.9 & 136.9 & 34\\ 
7 & TD & 7 & 32.8 & 139.0 & 34\\ 
8 & TD & 7 & 25.2 & 131.1 & 34\\ 
9 & TD & 7 & 22.7 & 122.0 & 34\\ 
10 & TD & 6 & 27.2 & 127.1 & 34\\ 
11 & TD & 6 & 20.1 & 114.8 & 34\\ 
12 & TD & 5 & 20.8 & 119.8 & 34\\ 
13 & TD & 4 & 18.7 & 114.0 & 34\\ 
14 & TD & 4 & 18.6 & 108.5 & 34\\ 
15 & TD & 6 & 23.2 & 122.4 & 31\\ \midrule
16 & DMD & 3 & 20.0 & 101.6 & 34\\ 
17 & DMD & 9 & 36.2 & 128.4 & 31\\ 
18 & DMD & 6 & 17.2 & 106.2 & 30\\ 
19 & DMD & 14 & 50.0 & 147.0 & 27\\ 
20 & DMD & 10 & 54.5 & 142.0 & 24\\ 
21 & DMD & 10 & 31.7 & 131.5 & 24\\ 
22 & DMD & 5 & 22.9 & 111.8 & 24\\ 
23 & DMD & 12 & 40.0 & 129.0 & 22\\ 
24 & DMD & 11 & 67.7 & 145.0 & 20\\ 
25 & DMD & 15 & 63.7 & 153.3 & 15\\ 
26 & DMD & 11 & 37.5 & 125.0 & 15\\ 
27 & DMD & 8 & 30.7 & 133.0 & 13\\ 
28 & DMD & 7 & 28.5 & 120.4 & 12\\ 
29 & DMD & 16 & 46.7 & 130.1 & 9\\ 
30 & DMD & 5 & 18.2 & 102.5 & 8\\
\bottomrule
\end{tabularx}
\end{table}



\subsection{Equipment}

Acceleration data from each participant were sampled at a rate of 100  Hz using an iPhone 11 and our Walk4Me smartphone application~\cite{ramli2023walk4me}. {We utilized the iPhone 11, which is equipped with a 3$-$axis MEMS accelerometer from STMicroelectronics. It recorded g-force measurements within ±8 g along each axis, maintaining a data sampling rate of up to \mbox{100 Hz.}} The phones were securely attached at the waist~\cite{chen2018establishing} with an athletic-style elastic belt enclosure, positioned approximately at the level of the lumbosacral junction. The~raw accelerometer signal was synchronized with video recordings captured by a GoPro camera at a rate of 30 Hz. An~observer marked the events where a participant passed the start or end of the duration or distance assigned to each activity using the web portal of the Walk4Me~system.

\subsection{Gait and Events Detection and Data~Analysis}

We collected the raw accelerometer signal from 30 participants, which included the \emph{\textcolor{black}{x}
}, \emph{\textcolor{black}{y}}, and~\emph{\textcolor{black}{z}} \textcolor{black}{axes} 
 (vertical, mediolateral, and~anteroposterior), along with the corresponding timestamps. Based on the findings of Zijlstra~\cite{zijlstra2003assessment}, we observed that the IC events were more distinguishable in the anteroposterior axis (\emph{\textcolor{black}{z}}$-$axis) compared to the other axes. Therefore, we used the anteroposterior signal from the raw accelerometer data to develop our method for counting the number of steps, estimating step length, and~calculating the total distance individuals walked at different~speeds.

\subsubsection{Method of Step~Detection}
\label{Method_of_Step_Detection}

Figure~\ref{Fig-X}A presents a raw accelerometer signal of the anteroposterior movement (\emph{\textcolor{black}{z}}$-$axis) from a TD participant during fast-walk speed calibration (SC-L4) {for 2.4 s}. The~steps in the anteroposterior signal are characterized by long wavelengths (low frequency), while other wavelengths (high frequency) represent noise signals. To~extract the steps, a~trained clinical evaluator reviewed testing videos and applied a low-pass filter~\cite{taborri2016gait} to individual participants’ data at each speed from a slow walk to a jog/run (range 0.5 Hz to 60 Hz) to smooth the signal and remove short-term fluctuations while preserving the longer-term trend indicating each step (Figure~\ref{Fig-X}A,B). 

We then identified the peak values of the filtered signal as the peaks occur only once per step in the filtered signal (Figure~\ref{Fig-X}C). The~number of peaks corresponds to the number of steps taken by the participant. Figure~\ref{SL}A shows the estimated number of steps using our method as blue dots, compared to the ground truth represented by a black line. The~built-in pedometer steps estimation is shown in~red.

\begin{figure}[H]

\begin{adjustwidth}{-\extralength}{0cm}
\centering 
\includegraphics[scale=0.98]{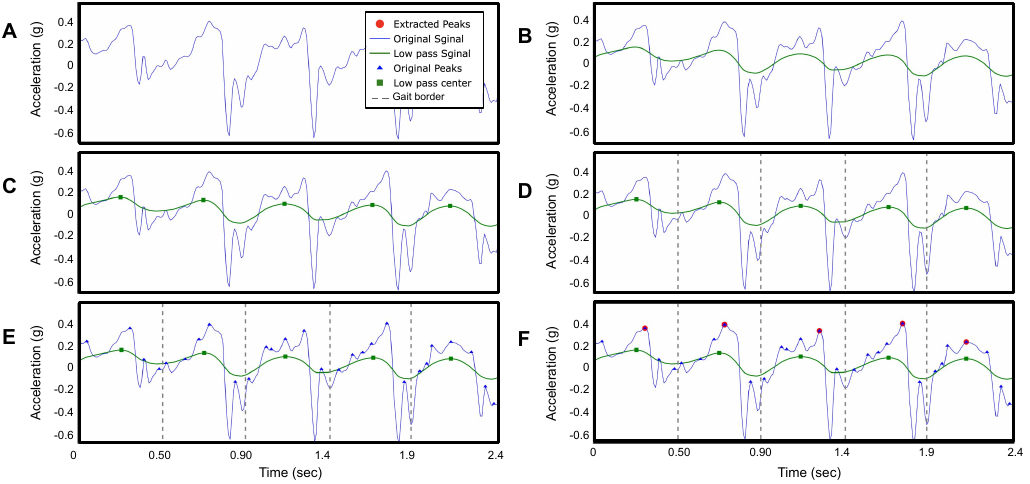}
\end{adjustwidth}
\caption{
\textcolor{black}{This} 
 figure presents the signal processing of the raw accelerometer signal of the anteroposterior movement (\emph{\textcolor{black}{z}}$-$axis) of participant ID 2 on fast-walk speed calibration (SC-L4) {for 2.4 s}. \linebreak  (\textbf{A}) Original raw accelerometer signal. (\textbf{B}) Filtered signal. (\textbf{C}) Peak detection of the filtered signal. \linebreak  (\textbf{D}) Locate the beginning and the end of each step. (\textbf{E}) Peaks detection of the original signal. \linebreak  (\textbf{F}) Locate the highest peak in the original signal.}\label{Fig-X}
\end{figure}
\vspace{-10pt}

\begin{figure}[H]
\includegraphics[scale=.26]{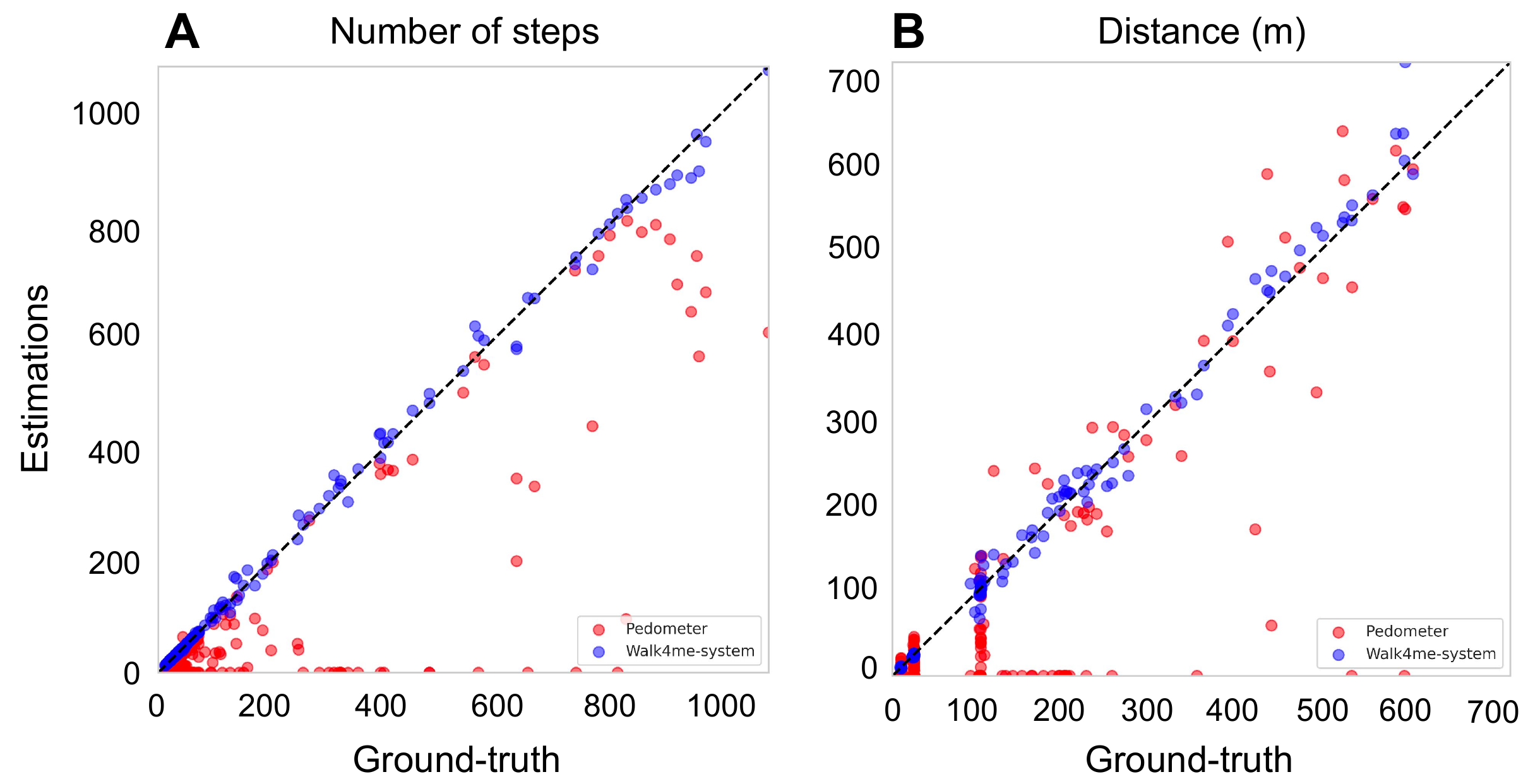}
\caption{{\em Cont.}} 
\end{figure}

\begin{figure}[H]\ContinuedFloat
\includegraphics[scale=.26]{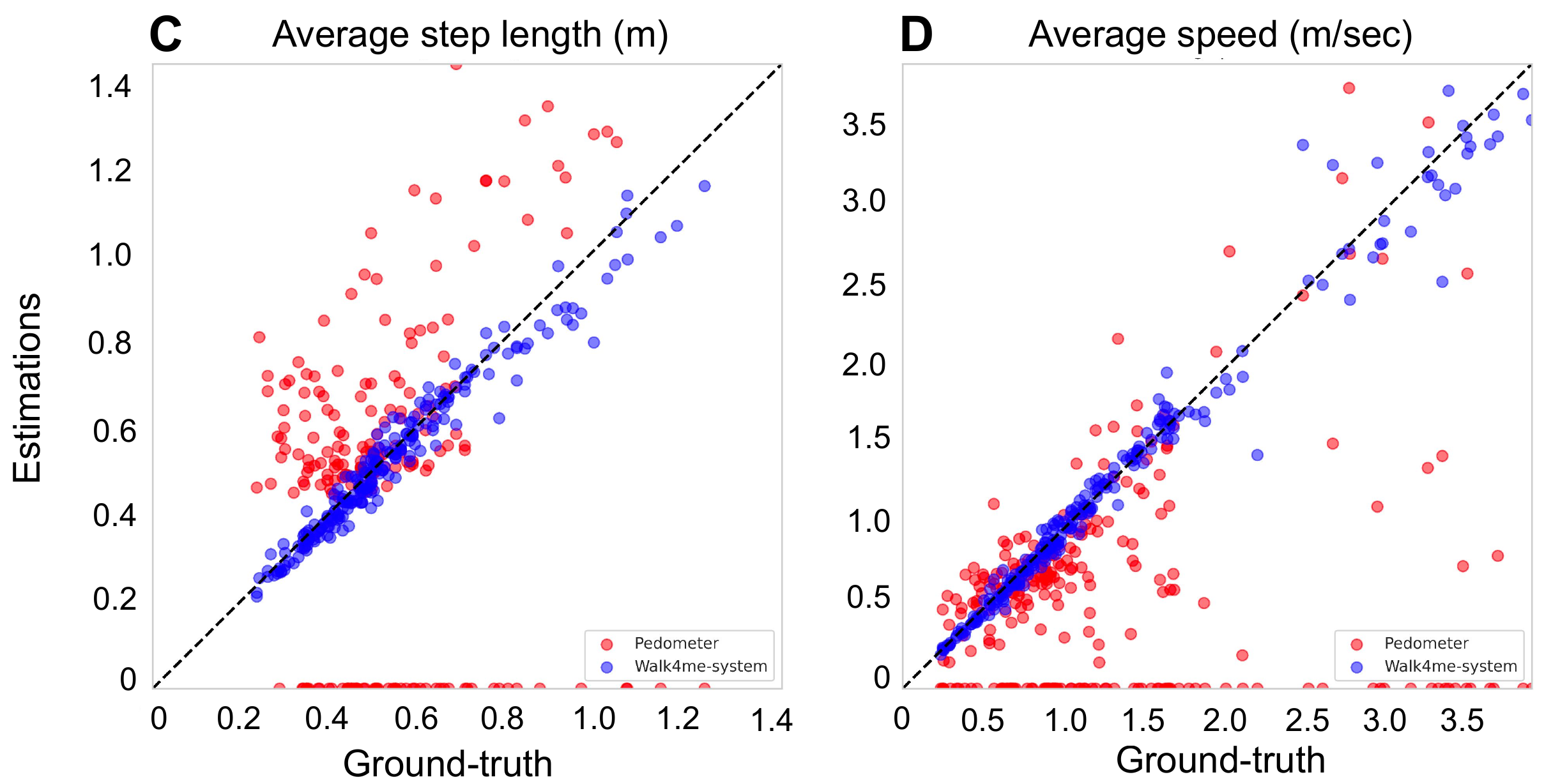}
\caption{\textcolor{black}{Comparison} 
 between the estimates and ground-truth values for both pedometers and our Walk4Me system to illustrate the accuracy across four key metrics: (\textbf{A}) the number of steps. The~adjusted R-squared of our Walk4Me is 0.9973, while the pedometer is 0.6826. (\textbf{B}) The distance in meters. The~adjusted R-squared of our Walk4Me is 0.9937, while the pedometer is 0.6987. (\textbf{C}) The average step length in meters. The~adjusted R-squared of our Walk4Me is 0.9595, while the pedometer is~\colorbox{white}{0.0094. (D) the average speed in meters per second.} 
}\label{SL}  
\end{figure}

\subsubsection{Method of IC~Detection}
\label{Method_of_IC_Detection}

To detect the IC events, we find the midpoint between two peaks in the filtered signal (Figure~\ref{Fig-X}D), which corresponds to the toe-off (TO) events during the gait cycle based on observation. We then identify all the peaks that occur within each step duration in the original acceleration signal (Figure~\ref{Fig-X}E). Next, we determine the maximum peak value (anteroposterior G), which corresponds to the time point of each IC (Figure~\ref{Fig-X}F).

\subsubsection{Method of Step Length Estimation Using~Regression}
\label{ABC}
We create an individualized {nonlinear} regression model~\cite{smyth2002nonlinear} for each participant to associate average peak acceleration values with step lengths. Figure~\ref{reg-pattren_REG_PAPER}A depicts the data flow of our model training and prediction process. Each model is trained using five different participant-selected calibration speeds (SC-L1 to SC-L5). For~each speed, we calculate the average acceleration peak values by taking the mean of all the peaks as described in Section~\ref{Method_of_IC_Detection}. To~calculate the average step length for training, we divide the observed ground-truth distance by the number of steps obtained from Section~\ref{Method_of_Step_Detection}. This process is repeated for each of the five calibration speeds (e.g., point SC-L4 in Figure~\ref{reg-pattren_REG_PAPER}A). The~resulting individualized equation through all five points allows us to input the peak acceleration value of any step within the participant's range of ambulatory velocity to estimate that step's length (shown as the green line in Figure~\ref{reg-pattren_REG_PAPER}A).

\subsubsection{Estimating the~Distance}
\label{Calculating_the_Distance}

After establishing the individualized model, it can be used on unseen data. We calculate the step lengths of all identified steps from a previously unseen event and accumulate them to calculate the total distance traveled by the individual. In~this project, we used \mbox{100 MRW,} 6 MWT, and~FW as input signals during the inference stage, as~shown in \mbox{Figure~\ref{SL}B,} and~compared the calculated distances with the ground-truth observed distances and the device's internal~pedometer.

\subsubsection{Calculating the Average Step~Length}

During the inference stage, to~calculate the average step length of an individual, we divide the distance estimated from Section~\ref{Calculating_the_Distance} by the number of steps obtained from Section~\ref{Method_of_Step_Detection}. Figure~\ref{SL}C shows the estimated average step length using our ML model as blue dots, compared to the ground-truth average step length represented by a black line. The~red dots represent the average step length estimated by the built-in~pedometer.

\subsubsection{Error Percentage~Rates}

To compare observed ground-truth step counts, distance traveled, and~average step lengths with our model's estimates and the pedometer estimates native to the mobile devices, we employed two methods. First, we calculated the aggregated error for all estimates by determining an error percentage rate ($Error_{rate}$) using Equation~(\ref{eq}).

\begin{equation}
Error_{rate} = \left|\frac{ \displaystyle\sum_{i}^{n} |V_{c}-V_{o}|_i - \displaystyle\sum_{i}^{n} |V_{o}|_i}{\displaystyle\sum_{i}^{n} |V_{o}|_i}\right| \times 100
\label{eq}
\end{equation}

The $Error_{rate}$ is calculated by aggregating the residual values of all participants ($i$) for all activities. The~residual is the difference between the proposed methods ($V_{c}$) and the total number of ground-truth observations ($V_{o}$). Then, the~total aggregated is subtracted from the total ground truth and divided by the total ground truth. Table~\ref{rate_table} compares the error percentage rate of step count, distance, and~average step length between our Walk4Me system and iPhone pedometer~measurements.

Second, to~evaluate the percentage error for each individual measurement and estimate pair, we subtracted the model estimate from the observed ground-truth measure and divided it by the ground-truth measure multiplied by 100 for each event. We computed mean (SD) percentage error for step count, distance traveled, and~step length parameters for calibration events SC-L1 to SC-L5 combined, and~separately for 6 MWT, 100 MRW, and~FW efforts combined, as~well as for all efforts combined. We compared the mean percentage error values between control participants and those with DMD using simple \textit{t}-tests for each~contrast.

\subsubsection{Gait Pattern~Representation}

\begin{figure}[H]
\includegraphics[scale=.27]{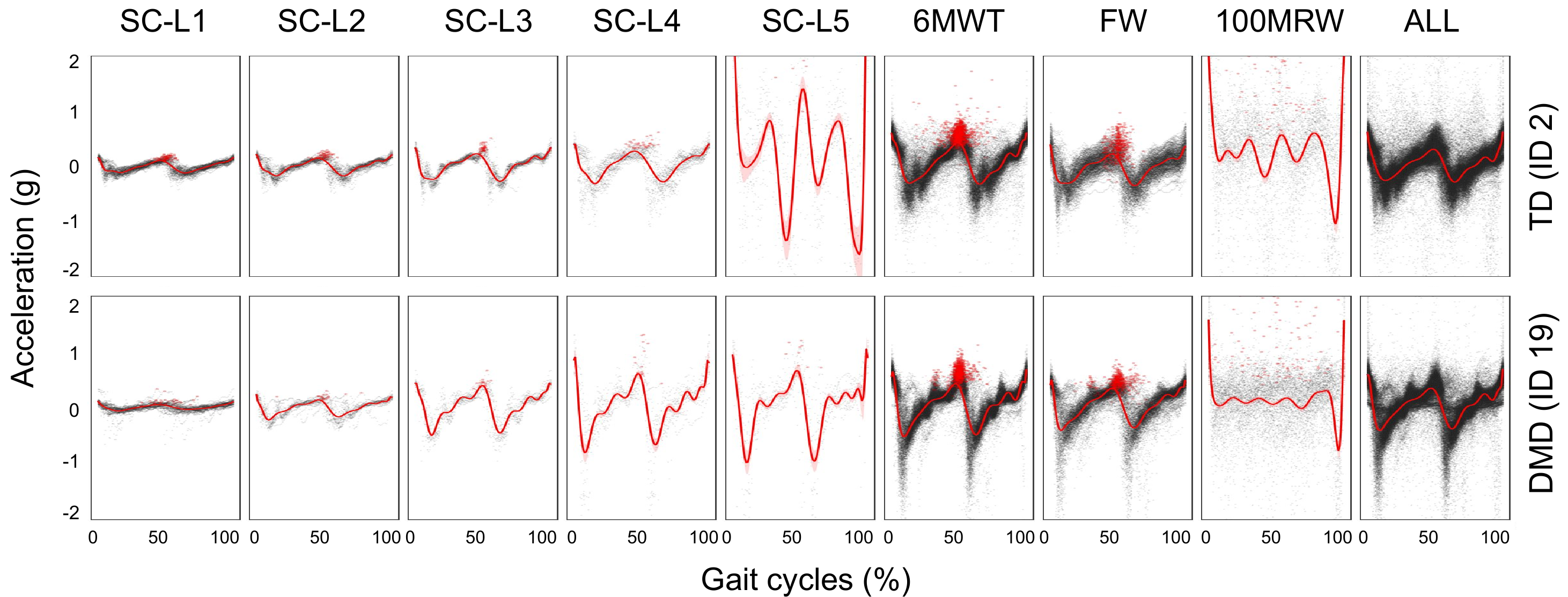}

\caption{\textcolor{black}{The} 
 gait pattern comparison of an individual with TD and an individual with DMD at varying speeds and a combination of multiple speeds and the effects of different speeds on gait~pattern.}\label{gait-pattren_REG_PAPER}\end{figure}

{After determining the boundaries between steps using the IC detection method discussed earlier, we generate a composite map of each step normalized to the gait cycle percentage. This allows for visual examination of the determined steps for irregularities or comparison of averaged accelerometer patterns between individuals (Figure~\ref{gait-pattren_REG_PAPER}). The~gait cycle is identified using peak detection at the IC event, marking the beginning and end of each step. The~average acceleration patterns are also calculated from all gait cycles across all activities and at various speeds. The~forward movement (\emph{\textcolor{black}{x}}$-$axis) is normalized to a time scale of 0 to 100\%. Using this method, we can identify the IC of every single step and estimate the step duration (Figure~\ref{Fig-X}F) without the need to use GRF~\cite{zijlstra2003assessment}. By~comparing the gait cycles of two participants (TD and DMD peers) at various speeds, distinctly different patterns of acceleration magnitude emerge (Figure~\ref{gait-pattren_REG_PAPER}), highlighting differences in the gait between the two participants.}

\section{Results}

In this study, we assessed the accuracy of step counts during walking, jogging, and~running using our Walk4Me system compared to the iPhone pedometer~\cite{bergman2012there}. We validated our results by comparing both systems with ground-truth data. Our findings, as~shown in Table~\ref{rate_table}, indicate that the Walk4Me system had an average step count error rate of 3.46\%, demonstrating reliable performance in accurately tracking steps at different speeds. The~combined error rates from participants with DMD and TD participants ranged from 1.26\% during slow walk pace (SC-L2) to 7.26\% during the fast 100 m run. In~contrast, the~iPhone's built-in pedometer showed an average error rate of 48.46\% during short- to moderate-distance tasks at varying gait velocities. The~iPhone pedometer had the lowest error rate of 36.35\% during the longer-duration fast-walk 6 MWT task, and~the highest error rate of 85.26\% during the short-duration jogging/running task~SC-L5.

\begin{table}[H]
\footnotesize

\caption{\textcolor{black}{The} 
 table compares the error percentage rate of the step count, distance, and~average step length between our Walk4Me system and iPhone pedometer~measurements.}\label{rate_table}

\begin{adjustwidth}{-\extralength}{0cm}
\resizebox{1.35\textwidth}{!}{%

\begin{tabular}{lllllllllll}
\toprule
~ & ~ & ~ & \multicolumn{3}{c}{ \textbf{Number of Steps}}  &  \multicolumn{3}{c}{\textbf{Total Distance}}  & \multicolumn{2}{c}{\textbf{Avg Step Length}}   \\ \midrule
~ & ~ & ~ & \textbf{GT *} & \textbf{Sys. **} & \textbf{Pedometer ***} & \textbf{GT *} & \textbf{Sys. **} & \textbf{Pedometer ***} & \textbf{Sys. **} & \textbf{Pedometer ***} \\
\textbf{Set} & \textbf{Activities} & \textbf{Case} & \textbf{(Steps)} & (\textbf{Error~\%)} & \textbf{(Error~\%)} & \textbf{(Meters)} & \textbf{(Error~\%)} & \textbf{(Error~\%)} & \textbf{(Error~\%)} & \textbf{(Error~\%)} \\
\midrule
\multirow{15}{*}{\makecell{\rotatebox[origin=r]{90}{Training set}}} & ~     &      TD &   976.0 &         1.64\% &     47.44\% &   359.30 &       5.37\% &   54.97\% &      6.43\% &  76.93\% \\
& SC-L1 &     DMD &   506.0 &         1.58\% &     57.31\% &   149.90 &       4.64\% &    65.7\% &      6.03\% &  75.07\% \\
&  ~     &     All &  1482.0 &         1.62\% &     50.81\% &   509.20 &       5.16\% &   58.13\% &      6.25\% &   76.1\% \\ \cmidrule{2-11}

& ~     &      TD &   782.0 &         1.41\% &     40.15\% &   359.80 &       4.48\% &    42.2\% &      5.38\% &  36.22\% \\
& SC-L2 &     DMD &   410.0 &         0.98\% &     49.02\% &   149.55 &        6.6\% &   39.82\% &      6.28\% &  59.94\% \\
& ~     &     All &  1192.0 &         1.26\% &      43.2\% &   509.35 &        5.1\% &    41.5\% &      5.79\% &  46.88\% \\ \cmidrule{2-11}
& ~     &      TD &   670.0 &         1.64\% &      49.4\% &   358.00 &       3.29\% &   50.66\% &      3.03\% &  34.39\% \\
& SC-L3 &     DMD &   347.0 &         1.15\% &     51.01\% &   149.80 &       3.79\% &   45.11\% &      4.47\% &  32.27\% \\
& ~     &     All &  1017.0 &         1.47\% &     49.95\% &   507.80 &       3.44\% &   49.03\% &      3.68\% &  33.44\% \\ \cmidrule{2-11}
& ~     &      TD &   562.0 &         1.96\% &     63.88\% &   358.70 &       2.93\% &   62.57\% &      2.86\% &  44.15\% \\
& SC-L4 &     DMD &   305.0 &         1.97\% &     69.84\% &   150.10 &       3.82\% &   62.64\% &      4.91\% &  35.34\% \\
& ~     &     All &   867.0 &         1.96\% &     65.97\% &   508.80 &       3.19\% &   62.59\% &      3.76\% &  40.28\% \\ \cmidrule{2-11}
& ~     &      TD &   373.0 &         3.75\% &      88.2\% &   352.90 &       5.63\% &   87.49\% &      9.16\% &  34.32\% \\
& SC-L5 &     DMD &   326.0 &         1.84\% &      81.9\% &   178.90 &       6.38\% &   72.78\% &      7.74\% &  24.62\% \\
& ~     &     All &   699.0 &         2.86\% &     85.26\% &   531.80 &       5.88\% &   82.54\% &      8.64\% &  30.76\% \\ \midrule
\multirow{9}{*}{\makecell{\rotatebox[origin=r]{90}{Test set}}} 
 &  ~     &      TD & \textcolor{black}{10,934.0} 
 &         2.32\% &     35.44\% &  7745.50 &       5.02\% &   27.92\% &      4.87\% &  34.62\% \\
&  6 MWT &     DMD &  7478.0 &         3.81\% &     37.68\% &  4949.00 &       5.48\% &   26.63\% &      5.85\% &  48.73\% \\
&  ~     &     All & \textcolor{black}{18,412.0} &         2.93\% &     36.35\% & \textcolor{black}{12,694.50} &        5.2\% &   27.42\% &      5.24\% &  39.87\% \\ \cmidrule{2-11}
& ~     &      TD &  1729.0 &         8.68\% &     62.81\% &  1543.50 &      10.95\% &   57.67\% &      6.97\% &   55.9\% \\
& 100 MRW &     DMD &  1906.0 &         5.98\% &     56.24\% &  1134.00 &       3.38\% &   50.41\% &       5.0\% &  54.62\% \\
& ~     &     All &  3635.0 &         7.26\% &     59.37\% &  2677.50 &       7.74\% &   54.59\% &       6.4\% &  55.53\% \\ \cmidrule{2-11}
&    ~     &      TD &  4460.0 &         3.27\% &     60.09\% &  3232.50 &       4.96\% &   58.39\% &      3.11\% &  64.92\% \\
&    FW &     DMD &  3767.0 &         5.07\% &     72.87\% &  2408.50 &       9.58\% &   67.05\% &      8.02\% &  81.02\% \\
&    ~     &     All &  8227.0 &          4.1\% &     65.94\% &  5641.00 &       6.93\% &   62.09\% &      5.14\% &  71.57\% \\ \midrule
&   All &     All & \textcolor{black}{35,531.0} &         3.46\% &     48.46\% & \textcolor{black}{23,579.95} &       5.83\% &   42.23\% &       5.8\% &   46.4\% \\
\bottomrule
\end{tabular}
}
\end{adjustwidth}

\noindent{\footnotesize{${^{*}}$ Ground truth. ${^{**}}$ Walk4Me system using our method. ${^{***}}$ Built-in pedometer in iPhone.}}

\end{table}

For distance measurement, our Walk4Me system showed an average error rate of 5.83\%, with~the lowest error rate of 3.9\% during the fast-walk SC-L4 pace and the highest error rate of 7.74\% during the fast 100 m run. The~iPhone's built-in pedometer had an average error rate of 42.23\%, with~task-specific error ranging from 27.42\% during the 6 MWT to 82.54\% during the SC-L5 jogging/running~task.

For step length measurement, our Walk4Me system showed an average error rate of 5.80\%, with~the lowest error rate of 3.68\% at a comfortable walking pace (SC-L3) and the highest error rate of 8.64\% during the short-term jog/run SC-L5 task. The~iPhone's built-in pedometer demonstrated an average error rate of 46.40\%, which varied from 30.76\% during SC-L5 to 76.10\% during~SC-L1.

In contrast to overall aggregate accuracy, the~mean (SD) accuracy of model predictions for individual events compared to ground-truth observations for step counts, distance traveled, and~step lengths is presented in Table~\ref{x} and depicted in Figure~\ref{SL}A--C. The~predicted and observed values for all three parameters showed a strong correlation (Pearson's \mbox{r = $-$0.9929} to 0.9986, \emph{p} < 0.0001). The~estimates demonstrated a mean (SD) percentage error of 1.49\% (7.04\%) for step counts, 1.18\% (9.91\%) for distance traveled, and~0.37\% (7.52\%) for step length compared to ground-truth observations for the combined 6 MWT, 100 MRW, and~FW tasks. There were no statistically significant differences in mean error percentages between control participants and those with DMD (\textcolor{black}{data not shown}
).


\begin{table}[H]
\tablesize{\scriptsize}
\caption{\textcolor{black}{The} 
 table shows the percentage error (correlation, \emph{p}-value, means, and~SD) of the predicted and observed values for step counts, distance traveled, and~step length vs. ground-truth~observations.}
\label{x}
\begin{adjustwidth}{-\extralength}{0cm}

\begin{tabularx}{\fulllength}{lm{1cm}<{\centering}m{1cm}<{\centering}m{1.7cm}<{\centering}m{1.7cm}<{\centering}m{1.7cm}<{\centering}m{1.7cm}<{\centering}m{1.7cm}<{\centering}m{1.7cm}<{\centering}}
\toprule
 &  & & \multicolumn{6}{c}{\textbf{Percentage Error of Calculated vs. Observed}} \\ 

  \cmidrule{2-9}
  
  & \multicolumn{2}{c}{\textbf{Ground Truth}} & \multicolumn{2}{c}{\textbf{Calculated vs. Obs. Steps}} & \multicolumn{2}{c}{\textbf{Calculated vs. Obs. Distance}} & \multicolumn{2}{c}{\textbf{Calculated vs. Obs. Step Length}} \\ 
  \midrule

{\textbf{Activities}} &{\textbf{\# Act.}} & {\textbf{\mbox{\textls[-25]{\# Steps}}}} & {\textbf{Correlation \mbox{\textls[-25]{(\emph{p}-Value)}}}} & {\textbf{Mean (SD)}} & {\textbf{Correlation \mbox{\textls[-25]{(\emph{p}-Value)}}}} & {\textbf{Mean (SD)}} & {\textbf{Correlation \mbox{\textls[-25]{(\emph{p}-Value)}}}} & {\textbf{Mean (SD)}} \\ \midrule

{\textbf{\textcolor{black}{SC-L1} 
 to SC-L5}} & 150 & 5257 &{0.9986 \linebreak  (\emph{p} \textless{} 0.0001)} & {1.2\% (2.87\%)} & {0.9946 \linebreak (\emph{p} \textless{} 0.0001)} &{$-$3.16\% (4.81\%)} &{0.9929 \linebreak (\emph{p} \textless{} 0.0001)} & $-$4.23\% (5.34\%) \\ 

{\textbf{6 MWT, 100 MRW, FW}} & {69} & 30,274 & {0.9972\linebreak (\emph{p} \textless{} 0.0001)} &{1.49\% (7.04\%)} & {0.9933 \linebreak(\emph{p} \textless{} 0.0001)} &{1.18\% (9.91\%)} & {0.9652\linebreak (\emph{p} \textless{} 0.0001)} & 0.37\% (7.52\%) \\

{\textbf{All}} & {219} & 35,531 & 0.9987   \linebreak (\emph{p} \textless{} 0.0001) &{1.29\% (4.59\%)} & {0.9969 \linebreak(\emph{p} \textless{} 0.0001)} & {$-$1.59\% (7.37\%)} & {0.9796 \linebreak(\emph{p} \textless{} 0.0001)} & $-$2.78\% (6.46\%) \\ \bottomrule
\end{tabularx}%

\end{adjustwidth}
\noindent{\footnotesize{The level of significance was set at 0.05. 
}}

\end{table}


\section{Discussion}

The use of travel distance and step length as gait metrics is essential for clinical gait assessment in the community setting. However, accurately measuring step length traditionally requires a clinical facility or gait lab with a trained observer present during the assessment session. Clinical assessment methods are considered the most detailed and ideal, but~their availability may be limited due to factors such as facility availability, staff availability, difficulties with patient travel to assessment locations, or~public health restrictions such as those related to COVID-19. Additionally, clinical observation methods can be susceptible to human error, such as observer fatigue or distraction, as~well as instrument errors, failed video recordings, or~obstructed views, which can limit the utility of the collected data. An~alternative option to overcome these limitations and facilitate more frequent and convenient collection of gait data in the community setting is to use off-the-shelf technologies such as pedometers, which are commonly built into smartphones and widely used in sports. However, it is crucial to assess the reliability of these devices, particularly when used for clinical purposes. Therefore, we conducted experiments to clinically validate the reliability of using a pedometer and compared the results with those obtained by~observers.

We propose an ML-based signal processing method using our Walk4Me system, which can estimate step counts, distance traveled, and~step lengths with increased levels of accuracy. The~advantage of our method is that it requires less observed interaction, only necessitating a short duration of time for five speed-calibration tests. Our system can automatically estimate distance and step length without the need for human interaction. Some of the source code and a demo of this paper can be found \colorbox{white}{at (28 May 2023)} 
 \url{https://albara.ramli.net/research/ic} along with some additional~results.

\section{Conclusions}

This study introduces a novel signal processing and machine learning technique that accurately identifies steps and step length based on the individual's gait style. Our findings demonstrate that using a single accelerometer worn near the body's center of mass can be more accurate than a standard pedometer. Our method can be applied to both healthy individuals and those with muscle disorders without the need for ground reaction force (GRF) measurements. To~our knowledge, this is the first study to propose a method that extracts CFs from raw accelerometer data across the attainable range of gait speeds in healthy participants and those with muscle disease. On~average, our method of counting steps and estimating stride length and distance traveled performs well when applied to longer structured sub-maximal clinical testing efforts and free-roaming self-selected pace travel. In~these settings, our methods surpass the pedometer functions native to the mobile devices we use. This will allow us to extend basic elements of gait analysis to community settings using commonly available consumer-level~devices.

\vspace{6pt} 




\authorcontributions{A.A.R., Conceptualization, Methodology, Software, Formal analysis, Writing, Supervision, Validation, Visualization, Investigation, Data Curation; X.L., Writing---Review and Editing, Methodology, Supervision; K.B., Investigation, Data Curation, Writing---Review and Editing; C.C., Writing---Review and Editing, Methodology, Supervision; E.G., Investigation, Supervision, Writing---Review and Editing; L.B.K., Investigation, Data Curation, Writing---Review and Editing; A.L., Investigation, Data Curation, Writing---Review and Editing; A.N., Investigation, Data Curation, Methodology; C.O., Investigation, Data Curation; D.R., Investigation, Data Curation, Writing---Review and Editing; J.W., Investigation, Data Curation, Writing---Review and Editing; D.A., Conceptualization, Methodology, Software, Analysis; C.M.M., Conceptualization, Resources, Funding acquisition; E.K.H., Conceptualization, Methodology, Software, Formal analysis, Writing, Supervision, Funding acquisition, Investigation. All authors have read and agreed to the published version of the manuscript.}

\funding{\textcolor{black}{This} 
\textcolor{black}{This research was funded by the US Department of Defense (grant number W81XWH-17-1-0477), the Muscular Dystrophy Association (grant number 646805) and intamural pilot funds from the University of California Center for Information Technology Research in the Interest of Society (CITRIS) and the Banatao Institute.}}

\institutionalreview{\textcolor{black}{  }
\textcolor{black}{The study was conducted according to the guidelines of the Declaration of Helsinki, and approved by the Institutional Review Board (or Ethics Committee) of the University of California Davis (IRB$\#1305174$, March 18, 2019).}}

\informedconsent{\textcolor{black}{  }}
Written informed consent was obtained from each participant prior to the initiation of study procedures.

\dataavailability{\textcolor{black}{Due to the human subject and health information privacy nature of the data and our institutional regulations, we will share information upon presentation of evidence of IRB or ethics board review and completion of appropriate data transfer agreements. Requests for data access can be addressed to the corresponding \textcolor{black}{author}.}}




\acknowledgments{\textcolor{black}{We} 
 would like to thank the students of the UC Davis EEC193A/B Winter 2020 Senior Design Projects Team (Nikki Esguerra, Ivan Hernandez, Zehao Li, and~Jingxuan Shi) for their work piloting proof-of-concept methods for clinical feature extraction. }

\conflictsofinterest{\textcolor{black}{The} 
 authors declare no conflict of interest.
} 



\appendixtitles{no} 
\appendixstart
\appendix

\begin{adjustwidth}{-\extralength}{0cm}

\reftitle{\highlighting{References} 
}

\end{adjustwidth}
\end{document}